\documentclass[12pt]{article}
\usepackage{natbib,mathpazo,graphicx,amsmath,setspace,lastpage,doi}
\usepackage[a4paper,margin=2cm,nohead]{geometry}

\usepackage{xspace}
\newcommand{\um}{\ensuremath{\mu \text{m}}\xspace} 
\providecommand*{\ped}[1]{\ensuremath{_\mathrm{#1}}}
\newcommand{\dmin}{d\ped{min}\xspace}

\def\0{\hbox{\phantom{\footnotesize\rm 1}}}

\begin{document}


\newcommand{\thetitle}{Spatial constraints underlying the retinal mosaics
  of two types of horizontal cells in cat and macaque}
\title{\thetitle}

\author{Stephen J. Eglen$^{1,\dagger}$, James C. T. Wong $^1$}
\date{$ $Revision: 1.26 $ $ \today}

\maketitle

\noindent $^{1}$ 
Cambridge Computational Biology Institute\\
Department for Applied Mathematics and Theoretical Physics\\
University of Cambridge\\
Wilberforce Road, Cambridge CB3 0WA, UK

\vspace*{2mm}

\noindent $\dagger$
Corresponding author.\\
\noindent Phone: +44 (0) 1223 765761

\noindent Email: S.J.Eglen@damtp.cam.ac.uk



\vspace*{3mm}

\noindent Postprint of Visual Neuroscience (2008) 25:209--214.  
\doi{10.1017/S0952523808080176}




\subsection*{Abstract}
Most types of retinal neurons are spatially positioned in non-random
patterns, termed retinal mosaics.  Several developmental mechanisms
are thought to be important in the formation of these mosaics.  Most
evidence to date suggests that homotypic constraints within a type of
neuron are dominant, and that heterotypic interactions between
different types of neuron are rare.  In an analysis of macaque H1 and
H2 horizontal cell mosaics, \citet{wassle2000} suggested that the high
regularity index of the combined H1 and H2 mosaic might be caused by
heterotypic interactions during development.  Here we use computer
modelling to suggest that the high regularity index of the combined H1
and H2 mosaic is a by-product of the basic constraint that two neurons
cannot occupy the same space.  The spatial arrangement of type A and
type B horizontal cells in cat retina also follow this same principle.

\subsection*{Key words} horizontal cells, retinal mosaics, minimal
distance model.

\clearpage

\section*{Introduction}

A defining feature for a type of retinal neuron is whether all neurons
of that type tile the retina in non-random patterns, termed ``retinal
mosaics'' \citep{cook1998_nato}.  This definition can also help us,
together with other anatomical and physiological properties, determine
whether a group of neurons should be classified as one type, or
subdivided into several types.  For example, cat beta retinal ganglion
cells (RGCs) are classed into two types, the on-centre beta RGCs and
the off-centre beta RGCs, partly because the mosaic of either the on-
or off-centre neurons independently tiles the retina and each mosaic
is much more regular than the combined mosaic of all beta RGCs
\citep{wassle1981beta}.  Furthermore, both cross-correlation analysis
and modelling suggest that these two types of neuron are independent
of each other in respect of positioning, as well as physiological
function, and, hence, may develop independently
\citep{wassle1981beta,eglen_vn2005}.  By contrast, \citet{wassle2000}
reported that for another pair of neuronal types, the
H1 and H2 horizontal cells in macaque:

\begin{quote}
  One would expect the nearest-neighbor distance of the combined
  mosaic to be smaller than that of the individual mosaics.  The
  regularity index [defined in Methods, below], however, is
  comparable, suggesting that the H1 and H2 cells are not arrayed
  completely independently.  It is possible, that some interaction
  between their mosaics during retinal development creates this
  overall regularity.  \citep[p597]{wassle2000}
\end{quote}

In this report we use computer modelling to investigate whether the
high regularity index of the combined mosaic of H1 and H2 neurons is a
product of type-specific interactions between the two
types, or whether it can be accounted for simply by anatomical
constraints resulting from the two cell types occupying the same
layer.  To generalise this question slightly, and to evaluate more
experimental data, we will compare the spatial patterning of
horizontal cells in macaque with cat \citep{wassle1978_horiz}.

\begin{figure}[htbp]
  \centering
  {\includegraphics{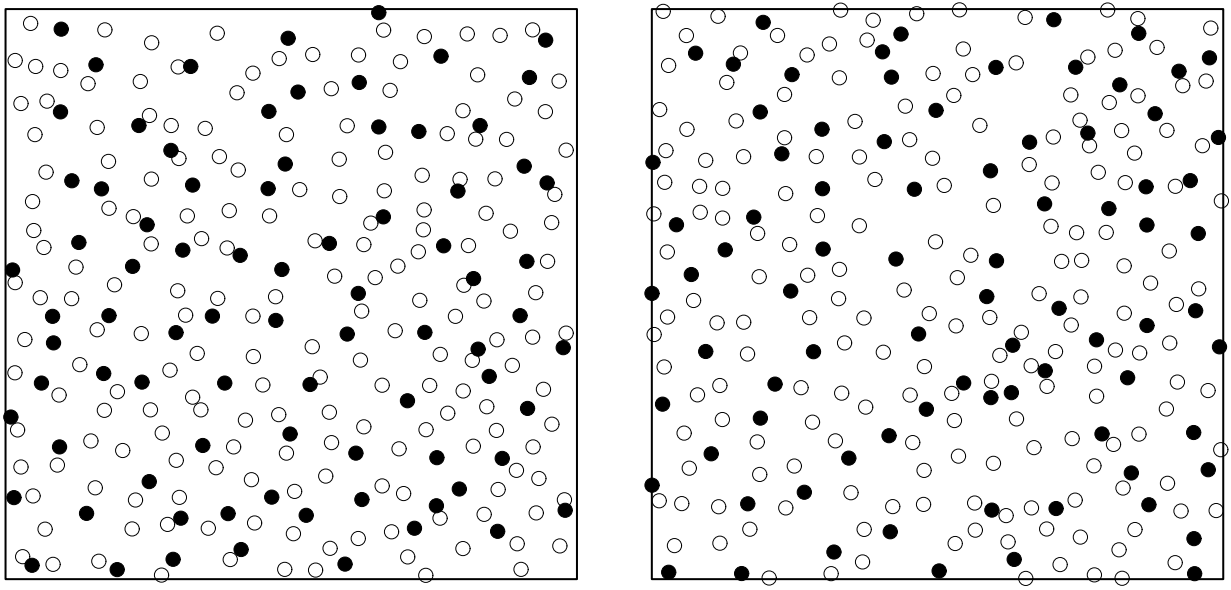}}
  \caption{Real and simulated horizontal cell mosaics.  Left: real
    mosaic (field A).  Right: example simulation.  Open circles denote
    H1 cells, filled circles denote H2 cells; cells drawn assuming
    10~\um diameter. Scale bar: 100~ \um. The
      simulated mosaic shows a close pair of H2 cells (halfway
      across, two-thirds up); such close pairs are rare but can occur
      when a homotypic exclusion zone is small.}
  \label{fig:mosaics}
\end{figure}

\section*{Methods}

\paragraph{Data sets}

Three horizontal cell fields were analysed: fields A and B are from
macaque (A: unpublished data; B: Figure 7 of \citet{wassle2000});
field C is from cat (Figure 12 of \citet{wassle1978_horiz}).
To keep our notation concise (rather than claiming any
equivalence of neuronal types across species), we denote type B cat
horizontal cells as ``type 1'', and type A horizontal cells as
``type 2'', in line with previously-noted similarities of primate H1
and other mammalian B cells \citep{lima2005}.  Fields were
digitised, and the cell location taken to be the centre of each soma.
Figure~\ref{fig:mosaics} shows an example real field along with a
matching simulation, defined next.

\paragraph{Bivariate \dmin model}

We have generalised the \dmin model
\citep{GalliResta1997} to simulate the positioning of two neuronal
populations within one field.  Each type of neuron has its own
homotypic exclusion zone ($d_1$ or $d_2$), but furthermore there is a
heterotypic exclusion zone ($d_{12}$) to potentially allow for
exclusions between the two types of neuron.  (The subscript $_{12}$
refers to an interaction between two types of neuron, whereas the
subscript $_{1+2}$ used below refers to all neurons irrespective of
type.)  First, we count the number of type 1 and type 2 neurons ($n_1$
and $n_2$), and simulate an area $A$ of the same size as the real field.
To initialise the simulation, we randomly position $n_1$ type 1
neurons and $n_2$ type 2 neurons within $A$.  Neurons are then
repositioned randomly within the field subject to two constraints:
that the nearest neighbour of the same type is greater
than some distance ($d_1$ for type 1 neurons, $d_2$ for type 2
neurons), and that the nearest neighbour of the opposite type is
greater than some distance $d_{12}$.  Each of the
distances $d_1, d_2, d_{12}$ is a random variable drawn from a Normal
distribution with a given mean and standard deviation.  Random values
lower than a lower limit (5 \um) are discarded, to prevent implausibly
small or negative \dmin values.  This birth and death process
\citep{ripley1977,eglen_vn2005} is repeated many times until
convergence (typically after each neuron has been moved ten times).

\paragraph{Null hypothesis}

Somata of both types of horizontal cell occupy the same stratum of the
inner nuclear layer (INL).  (In this study, we ignore the small
population of displaced horizontal cells that may be present in the
ganglion cell layer \citep{silveira1989,wassle2000}.)  Our null
hypothesis states that the developmental interactions between the two
types of neuron that influence their positioning are limited to
preventing somal overlap: any two neurons, regardless of type, cannot
come closer than some minimal distance.  In the context of our
simulations, this implies that the range of $d_{12}$ should match the
range of typical somal diameters of the two types of neuron.

\paragraph{Parameter estimates}

To fit one field, the free parameters in the model are the mean and
standard deviation of the three exclusion zones.  The homotypic
exclusion zones ($d_1$, $d_2$) were estimated first by fitting a
univariate \dmin model \citep{GalliResta1997} separately to the type 1
and type 2 neurons.  The size of the heterotypic exclusion zone,
$d_{12}$, was assumed to be of the same order as the soma diameter of
the horizontal cells, around 10~\um.  Parameters were fitted by
systematic searching over a range of plausible values.

\begin{table}
  \centering
  \begin{tabular}{rrrcrrr}
    \hline
    field 
    & $n_1$ & $n_2$ & width $\times$ height ($\um^2$)
    & $d_{1} (\um)$ 
    & $d_{2} (\um)$ 
    & $d_{12} (\um)$ 
    \\ \hline
    A     & 187 & 82 & $400 \0 \times \0402$
    & 22 $\pm$  \04 &  40 $\pm$  10 & 11 $\pm$ 3.0\\
    B     &  206 & 86 & $298 \0 \times \0300$
    & 21 $\pm$  \04 &  32 $\pm$ \08 & 12 $\pm$ 2.5\\
    C     & 300 & 85 & $723 \0 \times 1194$
    & 65 $\pm$   12 &  72 $\pm$ \08 & 14 $\pm$ 3.0\\
    \hline
  \end{tabular}
  \caption{Parameters for each \dmin simulation.  
    The number of type 1 and type 2 neurons ($n_1$, $n_2$) and the
    field size matches the values from the real field.  
    The diameter of each exclusion
    zone ($d_1$, $d_2$, $d_{12}$) is drawn from a Normal distribution,
    listed here as mean $\pm$ standard deviation.}
  \label{tab:params}
\end{table}

\paragraph{Assessing goodness of fit}

Two measures were used to quantitatively compare our model against the
real data, the regularity index (RI) and the K function.  The RI is
computed by measuring the distance of each non-border neuron to its
nearest-neighbour, and then dividing the mean of this distribution by
its standard deviation \citep{wassle1978}.  (Nearest-neighbour
distances of neurons at the border of a field are excluded as those
distances are unreliable.  A neuron is excluded if its Voronoi polygon
touches the boundary of the field.  This exclusion criterion accounts
for the small differences in RI between our work and those previously
reported.)  We measure three RI values: $RI_1$ (distance of each type
1 neuron to nearest type 1 neuron), $RI_2$: (like $RI_1$, but for type
2 neurons), and $RI_{1+2}$: (distance of each neuron to nearest other
neuron, irrespective of type).

For one population of neurons, $K(t)$ measures the number of cell
pairs within a given distance $t$ of each other
\citep{ripley1976,eglen_vn2005}.  For plotting purposes, we show $L(t)
= [K(t)/\pi]^{1/2}$.  This transformation discriminates between
exclusion ($L(t)<t$), clustering ($L(t)>t$) and complete spatial
randomness ($L(t)=t$).  We measure four L functions: $L_{1}$: pairs of
type 1 neurons; $L_{2}$: pairs of type 2 neurons; $L_{1+2}$: pairs of
neurons of either type.  Finally, $L_{12}$ measures the
cross-correlation, by constraining cell pairs such that one cell is
type 1 and the other is type 2.  Full details of these measures are
given elsewhere \citep{eglen2003,eglen_vn2005}.

To quantitatively evaluate the goodness of fit of the model to the
real data, each simulation was run 99 times with the same parameters,
but from different initial conditions.  Informally, if the measure
from the real data falls within the distribution of observed values
from the simulations, then the model fits the data.  This can be
quantified with a p value using a Monte Carlo ranking test.  A test
statistic ($T_i$) is measured for the K function of the real mosaic
($i=1$) and for each simulated mosaic ($i=2 \ldots 100$).  A p value
is then calculated by dividing the rank (smallest first) of $T_1$ by
100.  P values greater than 0.95 indicate a significant difference at
the 5\% level between model and data.  Full details of the test
statistic are given in \citet{eglen_vn2005}.

Computational modelling and analysis was performed in the R
environment, using the splancs package and Voronoi domain software
\citep{rcite,rowlingson1993,fortune1987}, as well as custom-written
routines.  The code is available from the authors upon request.

\section*{Results and Discussion}

\begin{figure}[htbp]
  \centering
  \vspace*{-1cm}
  {\includegraphics[width=15cm]{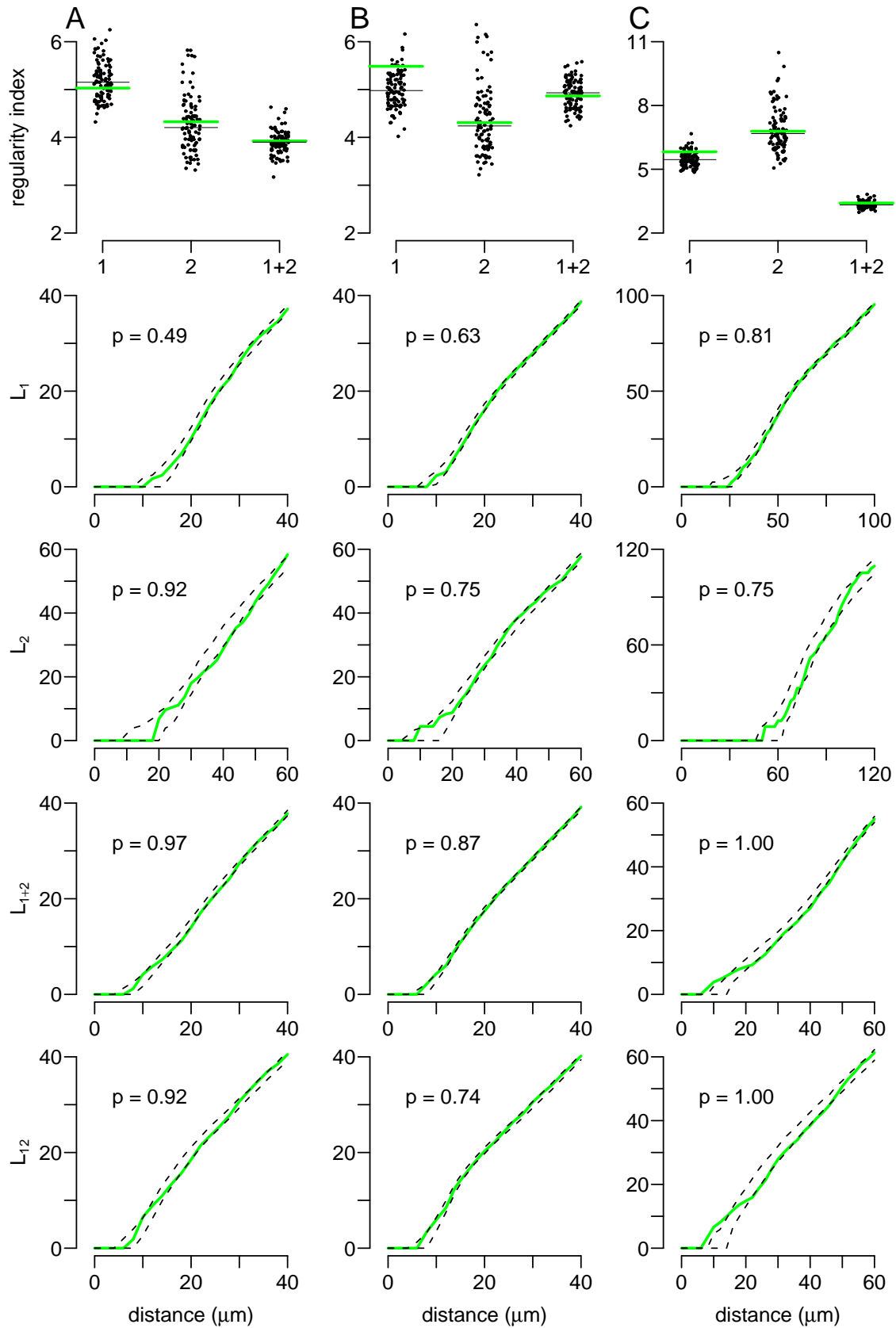}}
  \caption{(Colour online.)  Goodness of fit between data and model.
    Each column compares one field (A, B: macaque; C: cat) with
    simulations.  In
    the regularity index (RI) plots, the thick green line indicates
    the RI of the real data for either type 1, type 2, or all (1+2)
    neurons.  Black dots are RI values from 99 simulations, together
    with their median (thin black line).  For each of four L function
    plots for a field, the L function for the real mosaic is shown as
    a solid green line
    and dashed black lines indicate 95\% confidence intervals of
    simulations.  The p value is the goodness of fit between model and
    data.}
  \label{fig:gof}
\end{figure}

Table~\ref{tab:params} lists the parameters used for each bivariate
\dmin simulation.  The homotypic exclusion zones ($d_1$, $d_2$) were
independently fitted to each mosaic, whereas the heterotypic exclusion
zone ($d_{12}$) was set to just prevent neurons of opposite type from
occupying the same space in the inner nuclear layer.  (Mean values of
$d_{12}$ reported in Table~\ref{tab:params} are slightly higher than
estimates of somal diameter, suggesting that both cell bodies and some
initial portion of the primary dendrites contributed to steric
hindrance between neuronal types.)  Figure~\ref{fig:gof} shows that
for each field, the model generates mosaics that quantitatively match
the real mosaics, as assessed by both the RI and the L functions.  In
three (out of twelve) cases the goodness of fit p value is greater
than 0.95, indicating that formally there is a significant difference
between data and model.  Two of these cases concern both type 1 and
type 2 neurons from cat retina (field C).  Discrepancies in these two
cases are apparent over small distances (less than 10~\um between
neurons of opposite type), and may be due simply to difficulties in
reconstructing the position of pairs of opposite-type neurons that
seem to overlap in the field from the original publication
\citep[Figure 12 of][]{wassle1978_horiz}.  Small errors in determining
neuronal position are likely when considering the relative size of
individual neurons with the size of the sample field.  Overall,
however, the L functions for the data fit within the confidence
intervals of the model, suggesting that any disagreements between
model and data are quite small.

Regularity of a \dmin mosaic is influenced by both neuronal density
and the distribution of exclusion zone diameters.  For both macaque
fields, the median RI is higher for the simulated type 1 mosaics than
for the simulated type 2 mosaics; the opposite is true for the cat
field ($p < 0.001$ in each of three cases, Wilcoxon rank sum test).
The high RI of simulated cat type 2 mosaics, matching the observed
data, is due to the relatively low s.d. in the type 2 exclusion zone
(Table~\ref{tab:params}); if this is increased (e.g. from 8~\um to
16~\um), the median $RI_2$ decreases to 4.5 (data not shown), below
that of the type 1 mosaics.

\begin{figure}[htbp]
  \centering
  \includegraphics[width=15cm]{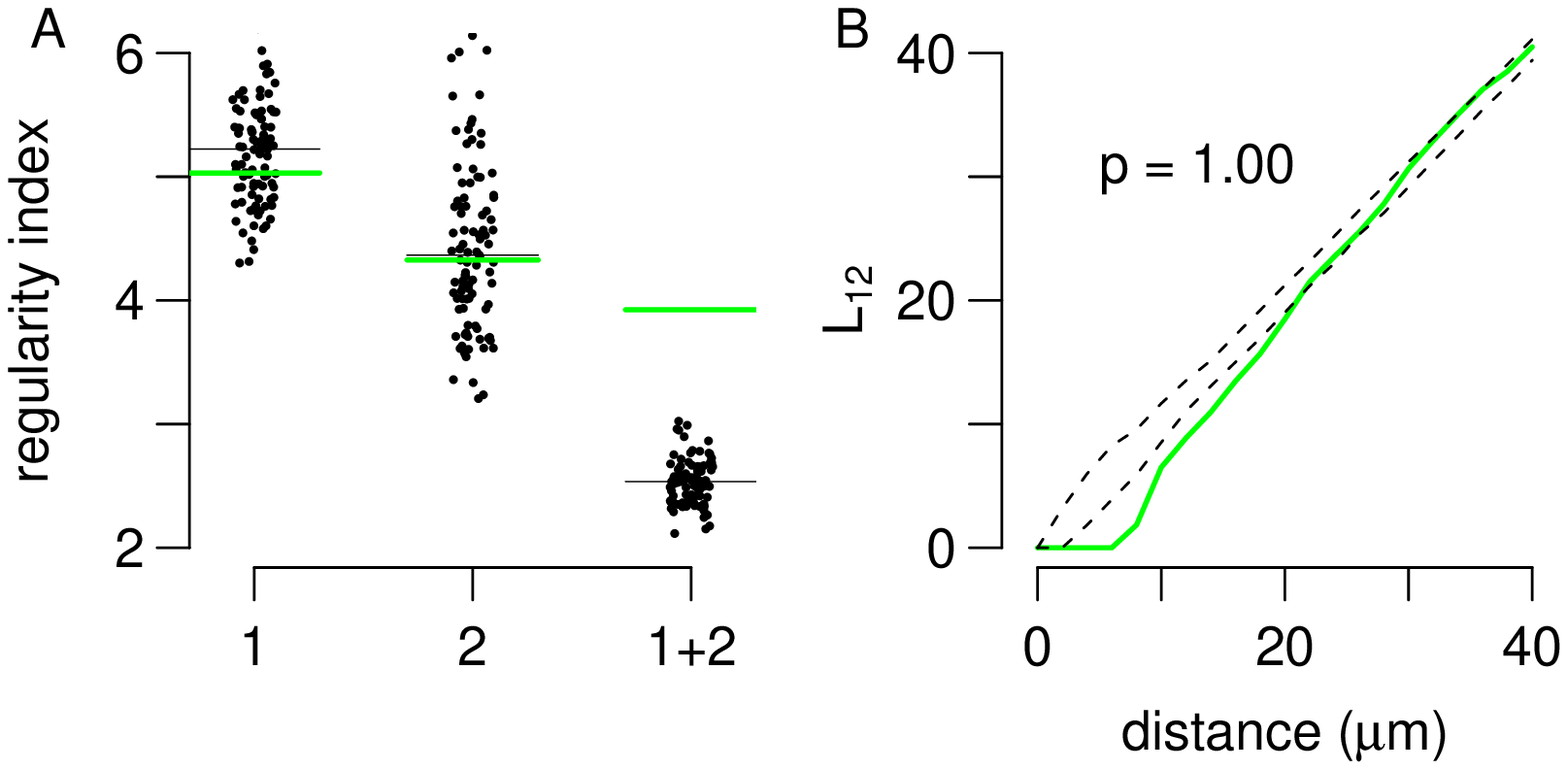}
  \caption{(Colour online.)  
    Results of bivariate \dmin simulation for
    field A with $d_{12} = 0$, and other parameters as listed in
    Table~\ref{tab:params}.  Results are presented as in
    Figure~\ref{fig:gof}.}
  \label{fig:nod12}
\end{figure}

The RI of the combined type 1 and 2 mosaic ($RI_{1+2}$) in each of our
fields is typically 3--5, matching the values observed experimentally.
At first glance, this might seem quite high, especially compared to a
theoretical expected value of around 1.9 for cells (of infinitesimally
small size) arranged randomly \citep{cook1996}.  Our model tells us
that this high RI is simply a by-product of superimposing two regular,
but independent, mosaics with somal exclusion.  This conclusion can be
supported in two ways.  First, by setting $d_{12}$ to zero, we
eliminate all heterotypic interactions.  (This allows for neurons of
opposite type to become arbitrarily close to one another, which is of
course not realistic, but allows us to specifically test the impact of
removing all heterotypic interactions.)  Figure~\ref{fig:nod12}A shows
that $RI_{1+2}$ drops considerably to a median of around 2.5, whereas
the fits to $RI_1$ and $RI_2$ remain good.  In this case, since there
is no positional constraint between opposite-type neurons, the L
function for the random simulations follows the theoretical
expectation $L_{12}(t) = t$.  The deviation between real data and
simulations is apparent up to at least 20~\um
(Figure~\ref{fig:nod12}B), as observed by the L function for the real
curve dropping well below the confidence intervals from the
simulations.

The second line of evidence to explain the high RI of the combined
(type 1 and type 2) mosaic is suggested by examining the fraction,
$f$, of the retinal area occupied by the cell bodies.  This can be
estimated by $f = ( (n_1 + n_2) \pi r^2) / {|A|}$ where $r=5~\um$ is
an estimate of radius of a horizontal cell soma, $n_1$ and $n_2$ are
the number of type 1 and type 2 horizontal cells, and $|A|$ is the
area of the field.  Table~\ref{tab:frac} shows that the fraction of
occupancy ($f$) correlates with the regularity of the combined mosaic
($RI_{1+2}$).  Furthermore, Table~\ref{tab:frac} also shows that the
mean number of trial cell positions rejected due to infringement of
the heterotypic constraint also correlates with the RI, even when
normalised for the number of potential pairwise heterotypic
interactions.  In this light, the cat and macaque mosaics are
generated by the same mechanism, and the lower regularity of the
combined mosaic in cat is due to the smaller effect of somal
exclusion.

\newcommand{\npairs}{\ensuremath{n\ped{pairs}}}

\begin{table}
  \centering
  \begin{tabular}{cccccc}
    \hline
    field & $RI_{1+2}$ & f & rejects & 
    $\npairs = n_1 \times n_2$ & rejects / \npairs
    
    \\ \hline
    C     & 3.4      & 0.04    & $\0242 \pm \010$ & 25500 & 0.01
    \\
    A     & 3.9      & 0.13    & $\0940 \pm \032$ & 15334 & 0.06
    \\
    B     & 4.9      & 0.26    & $7198 \pm 282$ & 17716 & 0.41
    \\
    \hline
  \end{tabular}
  \caption{Estimates of the fraction $f$ of sample retinal area
    occupied by all horizontal cell bodies in each field and incidence
    of heterotypic constraint enforcement.  Rows are
    sorted in order of increasing regularity index of the combined
    mosaic.  The mean ($\pm$ s.d.) number of times (per sweep)
    that the heterotypic constraint was broken, rejects, was counted
    over 99 simulations.  The final column shows the mean number of
    rejects divided by the number of pairs of opposite type neurons.}
  \label{tab:frac}
\end{table}

Our bivariate \dmin model is open to criticisms of biological
plausibility.  As previously noted, exclusion models show us that
local interactions are sufficient to generate regular patterns, but do
not inform us on how these interactions are mediated
\citep{GalliResta1997}.  Here we would suggest that the homotypic
exclusion zones are mediated by horizontal cell processes, perhaps
driving lateral migration during development \citep{reese1999}.  The
heterotypic interactions however are simply the result of steric
hindrance between cell bodies and primary dendrites.

In conclusion, results from our computational model suggest that the
high RIs observed for combined H1 and H2 mosaics in macaque are simply
a by-product of the two mosaics being positioned in the same stratum
of the INL; the mosaics may be developmentally independent in all
other respects.  This result agrees with our earlier work on beta RGCs
\citep{eglen_vn2005}, as well as other studies demonstrating a lack of
spatial correlations between many pairs of retinal neuronal types
\citep{mack2007,rockhill2000}.  Exceptions to this finding are rare
\citep{Kouyama1997,ahnelt2000}.

\paragraph*{Acknowledgements}

Thanks to Prof.\ Heinz W\"{a}ssle for providing the unpublished field,
labelled field A in this study, and to Prof.\ John Troy for critical
reading of this manuscript.  James Wong was supported by an EPSRC
studentship.


\bibliographystyle{jphysiol}
\bibliography{bihor}

\end{document}